\documentclass[preprintnumbers,amsmath,amssymb,prb,singlecolumn,superscriptaddress]{revtex4}

\usepackage{graphicx}   
\usepackage{dcolumn}    
\usepackage{bm}         

\usepackage{epsfig}

\begin{document}

\title{Tunable quantum dots in monolayer graphene}

\author{G. Giavaras}
\affiliation{Advanced Science Institute, RIKEN, Wako-shi, Saitama
351-0198, Japan}

\author{Franco Nori}
\affiliation{Advanced Science Institute, RIKEN, Wako-shi, Saitama
351-0198, Japan} \affiliation{Department of Physics, The
University of Michigan, Ann Arbor, MI 48109-1040, USA }



\begin{abstract}
We examine a graphene quantum dot formed by combining an electric
and a uniform magnetic field. The electric field creates a smooth
quantum well potential while the magnetic field induces an
exponential tail to the dot states. The states peak in the well
and the electrostatic barrier region as a result of the Klein
tunneling effect. Coupling between dot states which peak in
different regions can be achieved with the electric and magnetic
fields. The tunability of this dot with moderate external fields
could be used for designing quantum devices in monolayer graphene.
\end{abstract}

\maketitle

\section{Introduction}

The electronic properties of monolayer graphene are derived from a
two-dimensional Dirac equation. This makes graphene suitable to
explore quasi-relativistic effects in condensed matter physics as
well as attractive for quantum devices.~\cite{rozhkov} For these
two reasons various graphene-based devices are under
investigation. Lithographic confinement leading to quantum dot
behaviour has been achieved in nano-crystals of graphene with a
diameter of some tens of nanometers. Coulomb blockade,
single-electron transport, charge and spin spectroscopy have been
demonstrated in single and coupled
dots.~\cite{rozhkov,stampfer,molitor,guttinger}

Theoretical studies have shown that the electronic properties of
the dots formed in nano-crystals depend crucially on the type of
edges.~\cite{rozhkov} However, the edges cannot be routinely
controlled experimentally, because edge disorder seems to be
unavoidable and different types of edges may coexist in one
sample. Forming dots in monolayer graphene is necessary in order
to spatially isolate the dot states from the edges. In this case
the fabrication of the dot involves the application of external
fields. However, electrons in graphene are massless and via Klein
tunneling can penetrate any electrostatic potential
barrier.~\cite{rozhkov,katsnelson} Consequently, pure
electrostatic confinement in graphene is not
possible~\cite{matulis,hewageegana,chen,giavaras2009} and a
uniform magnetic field has to be applied.~\cite{giavaras2009} The
usual ease in tuning a dot electrically cannot be exploited in
graphene, though an exception may occur for zero-energy
states.~\cite{downing}

As shown in this work, the physics of a graphene dot formed by
combining an electric and a uniform magnetic field is rich and has
great potential for manipulation and control over the dot system.
The electric field creates a quantum well while the magnetic field
induces an exponential tail to the quantum states which is needed
for confinement. The electric field modifies the Landau level
spectrum by inducing energy levels within the Landau gaps. These
levels lie in a relatively low density of states. Here we
demonstrate that the tunability of these levels can be achieved
with both electric and magnetic fields.

The dot states exhibit features which arise due to the
relativistic character of graphene. When the magnetic field is
low, there is a class of states which peak in the quantum well
region but have also a large oscillatory amplitude in the
electrostatic barrier region due to Klein tunneling. As the field
increases, the amplitude in the barrier decreases exponentially
and the states become confined very close to the centre of the
dot. The suppression of the Klein tunneling with the magnetic
field may be employed in coupled dots. The interdot coupling is
expected to be strong (weak) in the low (high) magnetic field
regime. Strong interdot coupling seems to be possible even when
the dots are separated by a long distance. There is also another
class of states which peak mainly in the barrier region. These
states can couple to states which have a large amplitude in the
well, with a coupling strength that depends on the fields.
Numerical calculations show that it should be experimentally
possible to probe the dot properties in clean sheets of graphene.

The dot system studied here could allow the investigation of the
Klein tunnelling effect in a well-defined two-dimensional system.
In nano-crystals of graphene, the geometry is not circularly
symmetric and a spectral gap opening may prevent Klein tunnelling
from taking place. Moreover, the system that we study enables the
formation and coupling of multiple graphene dots in various
two-dimensional configurations. This may be more difficult to
achieve using nano-crystals. The proposed dot could even be formed
in suspended sheets of graphene in order to minimize interaction
and disorder effects due to the substrate.

In Sec.~\ref{dotsystem} we show that some of the basic properties
of the dot system can be extracted directly from the classical
energy-momentum relation of massless particles. We also present
the quantum mechanical model of the dot which is based on the
Dirac equation for the envelope functions of graphene.
Furthermore, in Sec.~\ref{dotsystem} a semi-analytical model is
solved for a piece-wise constant quantum well potential.
Calculations are presented in Sec.~\ref{exact}, where the
tunability of the dot with electric and magnetic fields is
demonstrated. The conclusions of this work are given in
Sec.~\ref{concl}.

\section{Graphene dot formed by electric and magnetic
fields}\label{dotsystem}

\subsection{Dot properties derived from the classical motion of massless particles}\label{clas}

Consider a graphene sheet in a vector potential $\textbf{A}=(0,
A_{\theta}, 0)$ and an electrostatic potential $V$, and for
simplicity choose $A_{\theta}$ and $V$ to have cylindrical
symmetry. Here $V$ defines a quantum well potential centered at
$r=0$ which is modelled as $V(r)=-V_{0}\exp(- r^2/d^2)$, with
$V_{0}\ge 0$. Also, $A_{\theta}$ generates a perpendicular
magnetic field to the graphene sheet which, in this work, is
chosen to be uniform, $\mathbf{B}=B\hat{z}={\bf\nabla} \times
\textbf{A}$, thus $A_{\theta}=Br/2$. The classical energy-momentum
relation of a massless particle moving in the fields $\textbf{A}$
and $V$, with energy $E$ and radial momentum $p$, is
\begin{equation}\label{enermoment}
v^2_{\text{F}}p^2 = (E-V)^2 - v^{2}_{\text{F}} \left(\frac{M}{r} +
e A_{\theta}\right)^2,
\end{equation}
where the term $M/r$ is due to the angular motion and
$v_{\text{F}}=10^{6}$ m s$^{-1}$ is the Fermi velocity in
graphene. The classical motion is restricted to the region where
$p^2>0$, and $p^2=0$ defines the classical turning points. The
ability of the magnetic field to induce confined states with an
exponential tail, regardless of momentum and energy, is based on
the observation that if $A_{\theta}\neq0$ then $p^2<0$ in the
asymptotic regime of large $r$ ($r\rightarrow\infty$). Therefore,
classical motion is not allowed in this regime. But if
$A_{\theta}=0$, then asymptotically $p^2>0$ and classical motion
is allowed, resulting in deconfined states with an oscillatory
tail. In this work we are interested in the energy range $-V_{0} <
E < 0$ and confined states which are needed for quantum dots,
henceforth $A_{\theta}\neq0$.

Equation~(\ref{enermoment}) shows that $p^2$ can take positive
values even within the barrier region before it eventually becomes
negative asymptotically.~\cite{baregion} For instance this can
happen when $|E-V_{0}|$ is large and $B$ is low. This regime can
always be arranged by tuning the electric and magnetic fields. If
$p^2>0$ in the barrier, then the quantum states are expected to
have an oscillatory spatial dependence; a property which is
related to the Klein tunneling effect for massless
particles.~\cite{katsnelson,calogerakos} In contrast, the states
of a particle with mass described by a Schr\"odinger equation
decay in the barrier.

\begin{figure}
\includegraphics[height=8.90cm,angle=270]{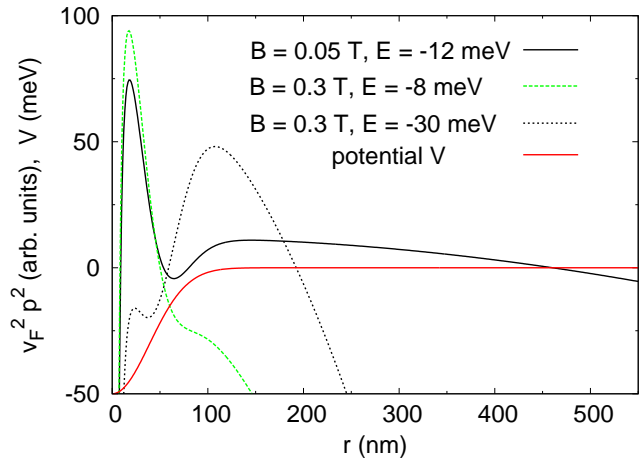}\\
\caption{(Color online) The figure shows $v^2_{\text{F}}p^{2}$,
defined in Eq.~(\ref{enermoment}), as a function of the radial
distance $r$ for three choices of magnetic field $B$ and energy
$E$. The potential well $V$ is also shown.}\label{states}
\end{figure}

Figure 1 shows $v^2_{\text{F}}p^2$ together with the potential
well. When $B=0.3$ T, then $p^2>0$ almost entirely in the barrier
(well) region for $E=-30$ meV ($E=-8$ meV), and hence the state is
expected to be confined in the barrier (well). When $B=0.05$ T and
$E=-12$ meV, then $p^2>0$ both in the well and barrier regions. In
this case the state can have a large amplitude both in the well
and the barrier. It can also be derived from
Eq.~(\ref{enermoment}) that if $B$ is high enough and/or the
angular momentum term is large ($M\gg0$) then there exists an
energy range for which $p^2<0$ for all $r$. So classical motion is
not allowed, suggesting that quantum states do not exist in this
range. Therefore, a gap, proportional to $B$ and $M$, is formed in
the energy spectrum. Then as the depth of the potential well
increases the gap closes and states are formed in the well and/or
barrier depending on the choice of energy.

Equation~(\ref{enermoment}) provides some insight into the
properties of the quantum dot and shows that different classes of
states can be defined according to the region which the states
tend to occupy. It also demonstrates the ability to tune at will
the dot states with electric and magnetic fields. However, it
cannot predict the energy levels of the dot and the exact pattern
of the states. For this reason we present below the exact model of
the quantum dot and identify its connection to the classical
energy-momentum relation in Eq.~(\ref{enermoment}).

\subsection{Quantum mechanical dot model}\label{quantum}

In monolayer graphene there are two valleys that have to be
considered, at the Dirac points $K$ and $K'$, respectively, of the
Brillouin zone.~\cite{rozhkov} In this work we assume that the two
valleys are uncoupled, as it has been demonstrated in many
experimental studies.~\cite{rozhkov} The physics is the same for
both valleys, thus we examine the dot properties only for one
valley. In the continuum approximation and for energies near the
Dirac points, the two-component envelope functions $\Psi$ satisfy
a two-dimensional Dirac equation
\begin{equation}
[v_{\text{F}}\bm{\sigma} \cdot ({\bf p} + e {\bf A})+V]\Psi =E
\Psi,
\end{equation}
where $\bm{\sigma}$ are the Pauli matrices, ${\bf p}$ is the
momentum operator, and $E$ is the energy. For a circularly
symmetric system, defined by the vector potential $\textbf{A}=(0,
A_{\theta}, 0)$ and the quantum well potential $V$ of the previous
subsection, we can write for the wavefunctions
\begin{equation}
\Psi =\frac{1}{\sqrt{r}}\left(\begin{matrix}
f_{1}(r)\exp[i(m-1)\theta]
\\
i f_{2}(r)\exp(i m\theta)
\end{matrix}\right),
\end{equation}
where $m=0,\pm1, ...$ is the angular momentum number and $r$,
$\theta$ is the radial distance and the azimuthal angle
respectively. The radial functions satisfy the
equations~\cite{giavaras2011}
\begin{subequations}
\begin{eqnarray}
Vf_{1} +\left(U+\gamma\frac{d}{dr} \right)f_{2} &=&Ef_{1},\label{radialequations1}    \\
\left(U-\gamma\frac{d}{dr}\right)f_{1}+Vf_{2}
&=&Ef_{2},\label{radialequations2}
\end{eqnarray}
\end{subequations}
with $v_{\text{F}}=\gamma/\hbar$ and
\begin{equation}
U=\gamma\frac{2m-1}{2r} + \gamma \frac{eA_{\theta}}{\hbar}.
\end{equation}
This eigenvalue problem can be diagonalized numerically to give
the eigenenergy $E$ and the corresponding two-component eigenstate
($f_{1}$, $f_{2}$). Here $f_{i}$ gives the probability amplitude
of finding an electron on one of the two sublattices of graphene.

Following the same method as that developed in
Ref.~\onlinecite{giavaras2009}, it can be shown that the state
$f_{i}$ satisfies the second order differential equation
\begin{equation}\label{single}
\frac{d^2f_{i}}{dr^2}+s(r)\frac{df_{i}}{dr}+w_{i}(r)f_{i}=0,
\end{equation}
with $s=V'/(E-V)$ and the prime denotes differentiation with
respect to $r$. Also
\begin{equation}
w_{i}=-\frac{U^2}{\gamma^2}\pm\frac{U'}{\gamma}\pm
s\frac{U}{\gamma}+\frac{(E-V)^2}{\gamma^2},
\end{equation}
where the minus (plus) sign is for $i=1$ ($i=2$). To derive a more
familiar Schr\"odinger-like equation we eliminate the first
derivative term. This can be done by writing $f_{i}$ in the form
\begin{equation}\label{f}
f_{i}=u_{i}\exp\left( - \frac{1}{2}\int s(r) dr \right),
\end{equation}
and substituting Eq.~(\ref{f}) into Eq.~(\ref{single}). This
procedure shows that the function $u_{i}$ is a solution to the
equation
\begin{equation}
\frac{d^2u_{i}}{dr^2}+k^{2}_{i}(r)u_{i}=0,
\end{equation}
with
\begin{equation}\label{k2}
k^{2}_{i}=\pm\frac{U'}{\gamma}\pm s\frac{U}{\gamma}
-\frac{s'}{2}-\frac{s^2}{4}+ \frac{v^2_{\text{F}}p^2}{\gamma^2},
\end{equation}
where the minus (plus) sign is for $i=1$ ($i=2$). Unlike the
coupled equations for $f_{i}$, the single equation for $u_{i}$ has
a more convenient form. The radial momentum $p^2$ in
Eq.~(\ref{enermoment}) can be directly identified in
Eq.~(\ref{k2}) with $M$ replaced by $(m-1/2)\hbar$. The quantum
model through Eq.~(9) suggests that the states are localised
within the region where $k^{2}_{i}>0$. This region is defined not
only by the $v^2_{\text{F}}p^2$ term but also by the additional
terms which appear in Eq.~(\ref{k2}). Confined states, i.e.,
states that decay asymptotically, occur when
$k^{2}_{i}(r\rightarrow\infty)\sim - v^2_{\text{F}} e^2
A^2_{\theta}<0$, which is satisfied for $A_{\theta}\neq0$. This
condition agrees with that derived from Eq.~(\ref{enermoment}).

One consequence of the Dirac equation is that the energy spectrum
of the graphene dot is unbound. However, depending on the external
fields, the energy levels and the quantum states can have simple
patterns. For instance, when $V=0$ and $m\ge1$ the energy spectrum
consists of two ladders (sets) of Landau levels (LLs) which can be
determined analytically.~\cite{clure} The ladders are separated by
a gap which is proportional to the field $B$ and angular number
$m$. The LLs are formed symmetrically with respect to $E=0$ and
the number of nodes in the corresponding Landau states increases
successively by one for each new level in each energy ladder. For
$m\le0$ one additional LL is formed at $E=0$ for which one of the
two components $f_{i}$ is zero.

The inclusion of a potential term $V\ne0$ modifies the spatial
region where the Landau states are localised. In addition energy
levels are formed within the Landau gaps. The simplest regime was
studied in detail in Ref.~\onlinecite{giavaras2010} within an
approximate model. It was shown there that for $m\ge1$ the dot
states are approximately localised in either of the two regions
defined by the two terms $E-V \pm U$, which create two
(independent) effective quantum wells. Each well contributes one
ladder of energy levels to the dot spectrum. For $m\le0$ the two
wells can no longer be defined because there is a point at which
the two curves $V\pm U$ cross. Nevertheless, exact numerical
calculations confirm the formation of the two ladders for $m\le0$
as well.~\cite{mneg}

The model developed in Ref.~\onlinecite{giavaras2010} is valid
provided the two effective wells have no common energy range. This
can be achieved when there is an energy gap between the minimum of
$E-V+U$ and the maximum of $E-V-U$. If this condition is not
satisfied the states of the two ladders can form a richer pattern.
Specifically, when $m$ is small the states of the upper ladder
tend to couple to the states of the lower ladder. This regime is
the main concern of the present work and it typically occurs when
the potential depth $V_{0}$ is large and the field $B$ is low. As
shown below, the induced coupled states can have a large amplitude
both near the centre of the quantum well and in the barrier
region. The coupling strength can be controlled at will by tuning
the external fields and hence different classes of states can be
probed. If $m$ is large the states are not affected by the quantum
well since they are localised in the asymptotic region where the
potential is constant. As a result the states for large $m$ behave
as Landau states.

The formation of the two ladders and the coupling between the
corresponding states have also been predicted theoretically in a
graphene dot system at zero magnetic field but with a spatially
modulated spectral (Dirac) gap in the energy
dispersion.~\cite{giavaras2011} In this system the states are
confined provided their energies lie inside the gap. The coupling
strength can be tuned with the potential depth and the induced
coupled states can be probed inside the spatial region where the
spectral gap is zero.~\cite{giavaras2011}

\subsection{Dot formed by a piecewise-constant potential well and uniform magnetic field}

\begin{figure}
\includegraphics[height=8.70cm,angle=270]{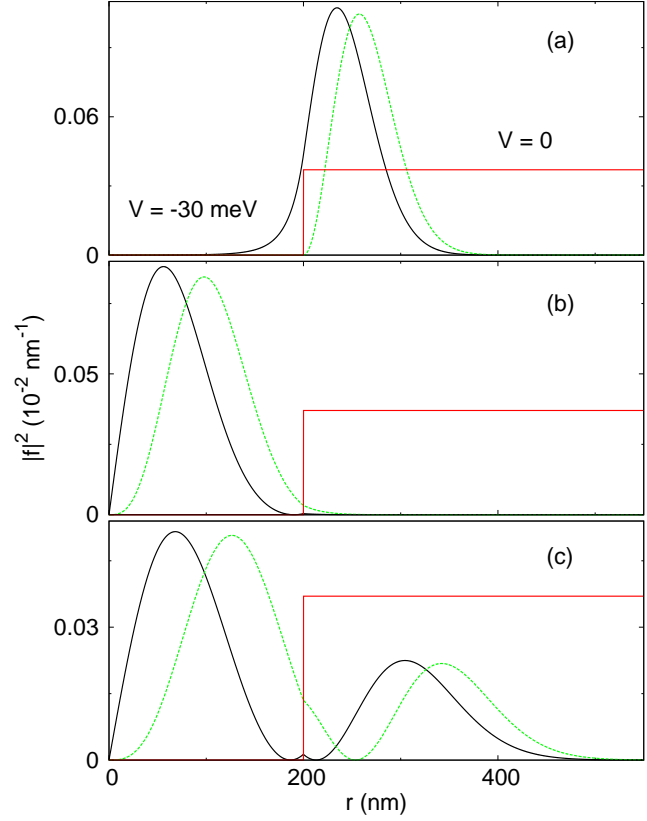}\\
\caption{(Color online) Quantum states for $m=+1$ and a
piecewise-constant potential well which has a radius $R=200$ nm
and depth $|v_{0}|=30$ meV. The magnetic field $B$ and the energy
$E$ are: (a) $B=0.2$ T, $E\approx-29.6$ meV, (b) $B=0.2$ T,
$E\approx-13.9$ meV, and (c) $B=0.09$ T, $E\approx-18.1$ meV.
Solid (dashed) curves show $|f_{1}|^{2}$
($|f_{2}|^{2}$).}\label{statesstep}
\end{figure}

This work is concerned with quantum dots formed in monolayer
graphene by combining electric and magnetic fields. In this
context, the simplest dot model that can be studied is when the
magnetic field is uniform and the induced potential well is
piecewise-constant, e.g., $V(r)=-|v_{0}|\Theta(r-R)$, where
$|v_{0}|$ is the depth of the well, $R$ is its radius and $\Theta$
is the step function. For completeness the method to solve this
problem is outlined here, though the reader can proceed directly
to Sec.~\ref{exact}. As in Sec.~\ref{quantum}
Eqs.~(\ref{radialequations1}) and~(\ref{radialequations2}) are
uncoupled and a second-order differential equation is derived for
each radial state $f_{i}$. Equation~(\ref{single}) for $s=0$ gives
\begin{equation}\label{deq1}
\frac{d^2f_{i}}{dr^2}+\frac{(E-V)^2}{\gamma^2}f_{i}-
\frac{U^2}{\gamma^2}f_i \pm \frac{1}{\gamma}\frac{d U}{dr}f_{i}=0.
\end{equation}
This equation is written in the more convenient form
\begin{equation}\label{conflu}
\frac{d^2f_{i}}{dr^2} - \frac{ a^2 r^4 + b_i r^2 + c_i }{r^2}
f_{i}=0,
\end{equation}
with the coefficients $a=eB/2\hbar$ and
\begin{subequations}\label{coeff}
\begin{eqnarray}
b_1=\frac{m eB}{\hbar}- \frac{(E-V)^2}{\gamma^2}, \quad c_1=
(m-1)^2-\frac{1}{4}, \\
b_2=\frac{(m-1)eB}{\hbar}- \frac{(E-V)^2}{\gamma^2}, \quad c_2=
m^2-\frac{1}{4}.
\end{eqnarray}
\end{subequations}
The form of Eq.~(\ref{conflu}) reveals that the states $f_{i}$ can
be expressed with the help of the confluent hypergeometric
functions $\mathcal{U}$ and $\mathcal{M}$
(Ref.~\onlinecite{abramowitz}). For example, when $m\ge1$ the
two-component state which is regular at the origin and decays
asymptotically has for $r\leq R$ the form
\begin{equation}\label{rlr0}
\left(%
\begin{array}{c}
  f_{1} \\
  f_{2} \\
\end{array}%
\right)=\eta e^{- ar^2/2} \left(%
\begin{array}{c}
  r^{d_{1}}
\mathcal{M}(A_1,B_1,ar^2)  \\
 c_0 r^{d_{2}}
\mathcal{M}(A_2,B_2,ar^2) \\
\end{array}%
\right),
\end{equation}
with
\begin{equation}
c_0=\frac{2\gamma a}{E+|v_{0}|}\left( 1 - \frac{A_{1}}{B_{1}}
\right).
\end{equation}
For $R\leq r$ the two-component state has the form
\begin{equation}\label{rlr0}
\left(%
\begin{array}{c}
  f_{1} \\
  f_{2} \\
\end{array}%
\right)=\beta e^{- ar^2/2}\left(%
\begin{array}{c}
 r^{d_{1}}
\mathcal{U}(A_1,B_1,ar^2)  \\
 g_0 r^{d_{2}}
\mathcal{U}(A_2,B_2,ar^2) \\
\end{array}%
\right),
\end{equation}
with
\begin{equation}
g_{0}=\frac{2\gamma a}{E}.
\end{equation}
The auxiliary parameters are
\begin{subequations}
\begin{eqnarray}
d_i=\frac{1+\sqrt{4c_i+1}}{2},
\\A_i=\frac{b_{i}}{4a}+\frac{1}{2}\left(\frac{1}{2} + d_{i}\right),
\\B_i=\frac{1}{2} + d_{i}.
\end{eqnarray}
\end{subequations}
The corresponding energies can be obtained from a standard
matching condition which requires that both $f_{1}$ and $f_{2}$ to
be continuous at $r=R$. This condition leads to an algebraic
equation for the energies which is solved numerically. Then having
calculated the energies and taking also into account the
normalization condition the ratio $\beta/\eta$ is determined. Some
quantum states for $m=+1$ are plotted in Fig.~\ref{statesstep}.
The region in which the states peak depends on the magnetic field
and energy. This behaviour is consistent with that derived from
the energy-momentum relation in Eq.~(\ref{enermoment}).

\section{Tuning the graphene dot with electric and magnetic
fields}\label{exact}

In any real sample of graphene a piecewise-constant potential well
cannot be generated. So a more realistic choice is made for the
potential well and, specifically, the potential well is chosen to
have a Gaussian form as in Sec.~\ref{clas} with $d=55$ nm.
Numerical calculations of the electrostatic potential in the
graphene sheet confirm that the Gaussian potential is a very good
approximation to the potential that is generated by gate
electrodes or a scanning tunneling microscope
tip.~\cite{giavaras2009} Also, the Gaussian potential produces a
smooth spatial variation on the scale of the lattice constant and
hence intervalley scattering can be ignored.

The results presented in this work are for the $m=+1$ states which
can peak very close to the centre of the potential well and thus
can be considered as quantum dot states. States with small values
of $m$ show similar trends to those with $m=+1$. As $m$ increases,
the states become localised in the asymptotic region where the
potential is constant and hence the effect of the quantum well on
the states becomes negligible. This property also occurs for a
repulsive potential profile.~\cite{park} Therefore, in the large
$m$ regime the corresponding quantum states are described
approximately by Landau states since only the vector potential and
angular momentum are important.

The dot states of interest have to be energetically isolated from
other states, that is, they have to lie in a region of low density
of states (DOS), in order to be detectable. If this condition is
satisfied then scanning tunneling microscopy (STM) could be used
to probe the local density of states and features of individual
quantum states. When there is no potential variation and a
perpendicular magnetic field is applied to a graphene sheet the
DOS is high at the LLs which form the energy spectrum~\cite{clure}
$\mathcal{E}_{N}=\pm\sqrt{2 e \hbar v^2_{F}B N}$, with
$N=n+(m+|m|)/2$, and $n$ being the radial quantum number. A
potential well induces energy levels within the Landau gaps where
the DOS can be low enough. As the external field is tuned, the dot
levels of interest have to lie within a specific Landau gap so
that their detection to be possible. For this reason we focus on
dot levels which lie between the first few LLs, for which the
Landau gaps are larger compared to those of highly excited LLs.

We assume that for the low magnetic fields considered in this work
there is no overlap between the LLs. Any LL broadening, for
example, due to disorder and interaction effects~\cite{yang,pound}
is small and it is typically less than $\sim 2$ meV. It may be
possible to achieve this condition in clean samples of graphene.
In particular, experiments in suspended sheets of graphene have
measured very large mobilities indicating that these sheets are
less sensitive to disorder and impurities introduced by the
substrate.~\cite{bolotin,du}

\subsection{Electric field effect on the dot}\label{electric}

As discussed in Sec.~\ref{quantum}, in the graphene dot system
formed by electric and magnetic fields, the energy levels for a
fixed value of $m$ can form two energy ladders separated by a
gap.~\cite{mneg} This configuration can be achieved for a high
magnetic field and a small well depth. When the well depth
increases, energy levels from the upper energy ladder fall in the
gap, while the corresponding states develop a large amplitude in
the well region. Eventually, when the well depth is large, states
which peak in the well couple to states of the lower energy
ladder, and this behaviour is manifested in the energy diagram by
the formation of anticrossing points.~\cite{coupling}

\begin{figure}
\includegraphics[height=8.65cm,angle=270]{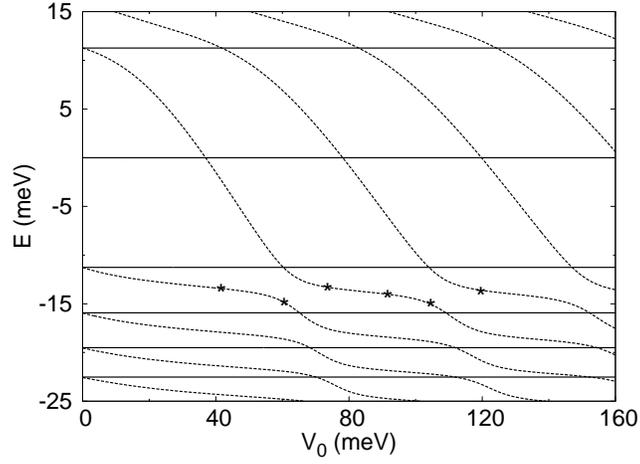}\\
\caption{Energy levels (dashed curves) as a function of the
potential depth $V_{0}$, for $m=+1$ and a magnetic field $B=0.1$
T. Some Landau levels are also shown (horizontal solid lines). The
quantum states for the marked energies are shown in
Fig.~\ref{statespot}.}\label{energypot}
\end{figure}

To demonstrate these effects, we show in Fig.~\ref{energypot} the
energy diagram as a function of the potential depth $V_{0}$ for a
magnetic field $B=0.1$ T. For $V_{0}=0$ the energy levels
correspond to the LLs of a graphene sheet. For a small $V_{0}$,
the energy levels which define the two energy ladders emerge into
the Landau gaps. When $V_{0}\sim0$ these levels are separated by a
gap which is approximately equal to the energy splitting between
the LLs $+1$ and $-1$ (Ref.~\onlinecite{llevel}). As $V_{0}$
increases, the gap closes and eventually energy levels of the
upper ladder anticross with energy levels of the lower ladder. The
general trend is that if the magnetic field is low the gap between
the two energy ladders is small and therefore the anticrossing
points appear for a small well depth $V_{0}$. For instance, the
first anticrossing point at $B=0.1$ T is formed for $V_{0}\approx
63$ meV (see Fig.~\ref{energypot}), while at $B=0.2$ T for
$V_{0}\approx 73$ meV. With a further increase of $V_{0}$, a state
of the lower energy ladder couples successively to states with
higher energy in the well, forming a series of anticrossing
points. Thus the required variation of $V_{0}$ to probe two
successive anticrossing points becomes smaller as the typical
energy splitting in the well decreases.

\begin{figure}
\includegraphics[height=8.8cm,angle=270]{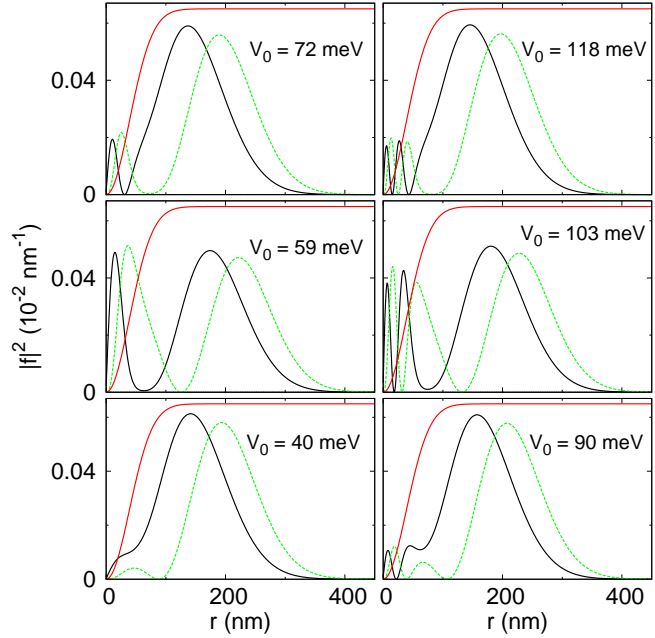}\\
\caption{(Color online) Quantum states for the marked energies
shown in Fig.~\ref{energypot}. The potential depth is $V_{0}$.
Solid (dashed) curves show $|f_{1}|^{2}$ ($|f_{2}|^{2}$). The
potential well (in arbitrary units) is also
shown.}\label{statespot}
\end{figure}

Here we focus on the state with the highest energy in the lower
ladder and its coupling to quantum well states. This state
corresponds for $V_{0}=0$ to the LL $-1$ and then for $V_{0}\ne0$
the coupling leads to a ``hybridised'' state with energy between
the LLs $-1$ and $-2$. Some examples of this hybridised state are
shown in Fig.~\ref{statespot}. Comparison of the states in the
left panels with those in the right panels shows that the number
of nodes for each radial component increases by one in the well
region~\cite{baregion} $r \lesssim 80$ nm, when a new anticrossing
point is formed in the energy diagram (Fig.~\ref{energypot}). This
happens because for each new anticrossing point a higher energy
state in the well is involved which typically has an extra node.
Near the anticrossing points the coupling is strong and hence the
amplitude of each state in the well region is large. This occurs,
for example, for $V_{0}=59$ meV and 103 meV. Away from the
anticrossing points the coupling is weak and the amplitude of the
states in the well decreases, e.g., for $V_{0}=40$ meV and 90 meV.

As seen in Fig.~\ref{statespot}, the hybridised state has a peak
in the barrier region~\cite{baregion} which, to a good
approximation, is insensitive to $V_{0}$. The reason is that
charge carriers in graphene are massless and they exhibit Klein
tunneling. This allows the states to develop an oscillatory
amplitude in the barrier. In conventional dots obeying
Schr\"odinger's equation the quantum states decay in the barrier.
Klein tunneling takes place in the two-dimensional graphene dot
when within a barrier region $k^{2}_{i}>0$ in Eq.~(\ref{k2}). This
condition can be arranged easily for the dot system by tuning the
external fields. Furthermore, the exact pattern of the state in
the barrier depends on the choice of energy. Specifically, the
number of peaks within the barrier increases for states which lie
between higher excited LLs with $\mathcal{E}_{N}<0$. However, the
narrowing of the corresponding Landau gaps in combination with the
expected broadening of the LLs in any real sample of graphene may
make the detection of these states difficult.

\begin{figure}
\includegraphics[height=8.65cm,angle=270]{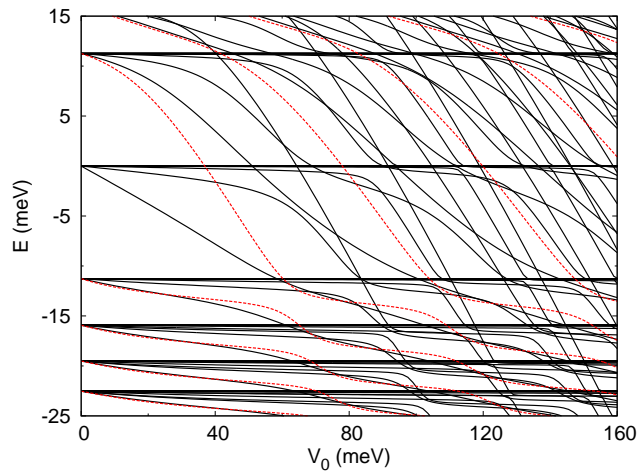}\\
\caption{(Color online) Energy diagram of the graphene dot as a
function of the potential depth $V_{0}$, for a magnetic field
$B=0.1$ T. The horizontal lines correspond to the Landau levels
which are independent of $V_{0}$. The dotted lines are the energy
levels for $m=+1$.}\label{dospot}
\end{figure}

Figure~\ref{dospot} illustrates the energy diagram of the dot as a
function of the potential depth $V_{0}$. Here all possible values
of $m$ are taken into account, but as explained above, for a large
$m$ the corresponding energy levels behave as LLs and thus they
have the dependence $\mathcal{E}_{N} \propto - \sqrt{BN}$. Between
the LLs there are regions where only a few levels lie, suggesting
that the corresponding states could be probed with STM
measurements.~\cite{miller} For this reason the energy of the
hybridised state in Fig.~\ref{statespot} is adjusted between the
LLs $-1$ and $-2$ where the DOS is relatively low. Inspection of
Fig.~\ref{dospot} shows that there are ranges of $V_{0}$ in which
the energy of this state is at least $0.5$ meV away from any other
energy levels. Thus it should be experimentally possible to probe
the pattern of the state in Fig.~\ref{statespot} in clean samples
of graphene with a small broadening of the LLs. The local DOS near
the centre of the dot is expected to show a low-high-low variation
as $V_{0}$ sweeps through an anticrossing point.

When the broadening of the LLs is large, the regions of low DOS
which occur between the LLs shrink making the detection of
individual dot states difficult; the states have to lie in a very
narrow energy range (window) in order to be detectable. A high
magnetic field $B$ may increase this energy range provided that
any broadening of the LLs increases much slower than the Landau
gaps. However, when $B$ is high a large $V_{0}$ is needed for
anticrossing points to appear and the states to follow the pattern
in Fig.~\ref{statespot}. In this case the disadvantage is that the
DOS between the LLs increases as $V_{0}$ and $B$ increase,
suggesting that the magnetic field cannot be chosen arbitrarily
high. Also, if in an experiment it is desirable for the state to
peak in the barrier region and far from the centre of the dot then
the magnetic field has to be kept low.

In order to deal with the broadening of the LLs it may be
necessary to probe hybridised states between the LLs $0$ and $-1$,
for which the Landau gap is the largest. Though, only states with
$m\le0$ can, for certain parameter regimes, form anticrossing
points at energies which lie between the LLs $0$ and $-1$. This
can be understood from the fact that for $V_{0}=0$ only states
with $m\le0$ contribute to the zeroth Landau level. Then as
$V_{0}$ increases, their energies emerge into the zeroth Landau
gap and eventually anticross with energy levels of the upper
ladder. States with $m\le0$ can also form anticrossing points in
excited Landau gaps as happens with the $m=+1$ states in
Fig.~\ref{energypot}.

\subsection{Magnetic field effect on the dot}\label{magnetic}

The effect of a magnetic field on the dot is now examined when the
potential well is fixed. In Sec.~\ref{electric} it was shown that,
when the electric field increases, states belonging to different
energy ladders become coupled. Moreover, the coupling is strong
near the anticrossing points. In this section it is demonstrated
that when the magnetic field increases the coupling is suppressed
resulting in states which correspond approximately to individual
states in each ladder.

\begin{figure}
\includegraphics[height=8.7cm,angle=270]{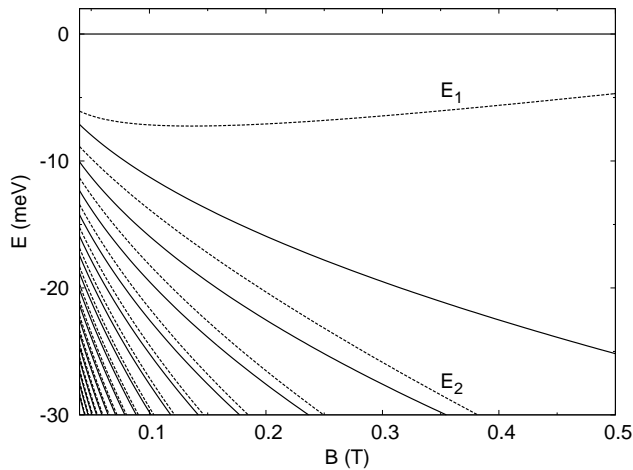}\\
\caption{Energy levels (dashed curves) as a function of the
magnetic field $B$, for $m=+1$ and a potential depth $V_{0}=51$
meV. Some Landau levels are also shown (solid curves). The quantum
states for the energy levels $E_{1,2}$ are shown in
Fig.~\ref{statesmag}.}\label{energymag}
\end{figure}

In Fig.~\ref{energymag} the energy diagram is plotted as a
function of the magnetic field when the depth of the potential
well is $V_{0}=51$ meV. We are interested in the two highest
energy levels below zero (for $m=+1$) which are denoted by $E_{1}$
and $E_{2}$. In the field range $0.04< B<0.5$ T, the $E_{1}$ level
lies between the LLs $0$ and $-1$, while for $0.04< B <0.35$ T the
$E_{2}$ level lies between the LLs $-1$ and $-2$. Thus quantum
states can be arranged to lie between successive LLs in a wide
field range. Calculations show that this effect is robust and can
be achieved for different well depths and $m$ values. In addition,
as shown below within a specific magnetic field range the spatial
region in which the states are localised changes drastically,
confirming that the tunability of the dot is feasible with a
uniform magnetic field.

\begin{figure}
\includegraphics[height=8.50cm,angle=270]{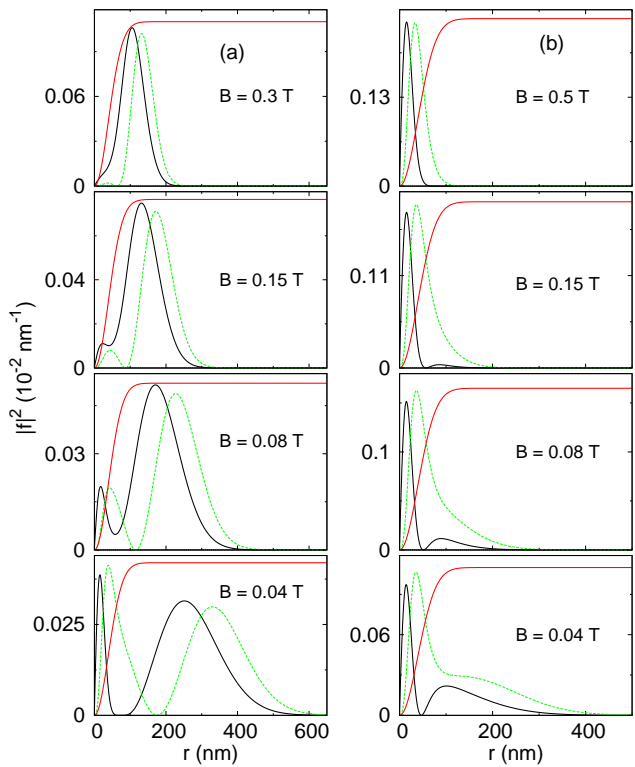}\\
\caption{(Color online) Quantum states for the energies shown in
Fig.~\ref{energymag}. For the energy level (a) $E_{2}$ and (b)
$E_{1}$. Solid (dashed) curves show $|f_{1}|^{2}$ ($|f_{2}|^{2}$).
The potential well (in arbitrary units) is also
shown.}\label{statesmag}
\end{figure}

Figure~\ref{statesmag}(a) shows the quantum state corresponding to
the energy level $E_{2}$ in Fig.~\ref{energymag}. This
``hybridised" state is formed due to the coupling between the
states of the two energy ladders. In particular, the peak in the
well (barrier) region~\cite{baregion} is due to the state in the
upper (lower) ladder. When $B=0.04$ T the coupling is strong and
the hybridised state peaks both in the well region, that is $r
\lesssim 80$ nm, and the barrier.~\cite{peakregion} As $B$
increases, the coupling is suppressed and the amplitude of the
hybridised state very close to the centre of the well decreases.
Consequently, the hybridised state takes approximately the form of
the state in the lower energy ladder. Also, when the field
increases the magnetic confinement becomes stronger, therefore the
state in the lower ladder shifts closer to the centre of the dot
and becomes localised in a narrower region. These trends are also
displayed by the hybridised state in Fig.~\ref{statesmag}(a),
which has a large peak in the region $200\lesssim r \lesssim400$
nm at $B=0.04$ T, whereas  the peak occurs around $50\lesssim r
\lesssim150$ nm at $B=0.3$ T.

The magnetic-field-dependent coupling is also responsible for the
behaviour of the ``hybridised" state corresponding to the energy
level $E_{1}$ [see Fig.~\ref{statesmag}(b)]. At $B=0.04$ T the
state peaks in the well region $r \lesssim 80$ nm, but has also a
large amplitude in the barrier region before it eventually decays
asymptotically. Klein tunneling is involved in the low field
regime but is not so pronounced because there is only a small
region with $k^{2}_{i}>0$ within the barrier. Then as $B$
increases the amplitude of the two components in the barrier is
gradually suppressed and the state becomes confined near the
centre of the well. For $B=0.5$ T the hybridised state has
approximately the form of the state in the upper energy ladder and
$k^{2}_{i}<0$ everywhere in the barrier. Thus no Klein tunneling
occurs at high $B$. The regime of high magnetic field as well as
the symmetry between a dot and an antidot system have been
investigated theoretically in Ref.~\onlinecite{maksym}. A graphene
dot system especially designed to probe the Klein tunneling effect
in a two-dimensional geometry has been studied in
Ref.~\onlinecite{giavaras2009}.

\begin{figure}
\includegraphics[height=8.70cm,angle=270]{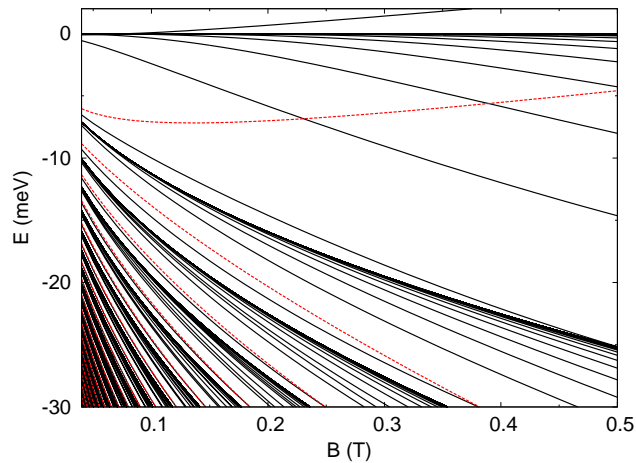}\\
\caption{(Color online) Energy diagram of the graphene dot as a
function of the magnetic field $B$, for a potential depth
$V_{0}=51$ meV. The high-density regions have the same magnetic
field dependence as that of the Landau levels in graphene
($\mathcal{E}_{N}\propto-\sqrt{BN}$). The dotted lines are the
energy levels for $m=+1$.}\label{dosmag}
\end{figure}

Figure~\ref{statesmag} demonstrates that the magnetic
field-induced suppression of the amplitude of the states in the
barrier is a strong effect. This property may be useful for a
system of two neighboring graphene dots separated by a long
distance. In this system the interdot coupling can be tuned with
the magnetic field instead of using gate electrodes as in
conventional dots. The coupling is expected to be strong in the
low field regime, provided Klein tunneling is involved. The
coupling can be strong even when the two dots are far apart, a
situation that cannot be arranged easily in conventional
semiconductor dots.

In Fig.~\ref{dosmag} we show the energy diagram of the dot as a
function of the magnetic field $B$. All values of $m$ are taken
into account, but as explained above, when $m$ is large the energy
levels have the LL dependence $\mathcal{E}_{N} \propto -
\sqrt{BN}$. As seen in Fig.~\ref{dosmag}, regions of very low DOS
are formed between the lowest LLs. This condition can be arranged
when the quantum well is narrow enough so that only a limited
number of $m$ values contributes energy levels between the LLs.
The energy level $E_{1}$ in Fig.~\ref{energymag} crosses some
other levels for a few values of $B$ but it is isolated from any
other levels in a wide field range. This property makes possible
the resolution of the corresponding state and its variation with
the magnetic field [Fig.~\ref{statesmag}(b)]. Investigation of the
energy diagram in Fig.~\ref{dosmag} shows that for $0.04 <B< 0.3$
T the energy level $E_{2}$ in Fig.~\ref{energymag} is at least
$0.4$ meV away from any other levels. This suggests that the
corresponding quantum state [Fig.~\ref{statesmag}(a)] could be
detected using STM measurements similar, for example, to those in
Ref.~\onlinecite{miller}.

To demonstrate the formation of confined states in a graphene
quantum dot the main focus of the experiment should be on the
state with energy level $E_{1}$ and in the field range $0.15
\lesssim B \lesssim 0.36$ T. The reason for this choice is
twofold. First, this state lies in a very low DOS region and thus
it could be resolved using standard charge sensing measurements.
Second, the energy of this state is away from the LLs and its
detection should be possible when the LL broadening is $\lesssim$
10 meV. This value has to be smaller when the magnetic field is
not in the above range or when the experiment probes the energy
level $E_{2}$.

\section{Discussion and Conclusion}\label{concl}

This work investigated the properties of a graphene quantum dot
formed by the combination of electric and magnetic fields. The
electric field creates a smooth quantum well potential which could
be generated using gate electrodes. Because of the Klein tunneling
effect in graphene the well cannot confine electrons. However,
when a uniform magnetic field is applied perpendicular to the
graphene sheet the states decay asymptotically as needed for
confinement. The electric field modifies the Landau level spectrum
by inducing levels within the Landau gaps. It was demonstrated
that the states which correspond to these levels are tunable with
the electric and magnetic fields. This property has also been
considered in one-dimensional systems of graphene such as
wave-guides and wires.~\cite{bliokh2009,bliokh2010}

Some of the basic properties of the dot system were qualitatively
extracted from the classical energy-momentum relation of massless
particles. These properties were then confirmed and quantified
with exact numerical calculations. It was found that some states
can peak both in the potential well and within the barrier because
of the Klein tunneling effect in graphene. This behaviour cannot
be observed in a non-relativistic dot described by a Schr\"odinger
equation. States which peak in the barrier region occur also in
one-dimensional systems of graphene and have been examined in
Refs.~\onlinecite{nori} and~\onlinecite{pereira}.

The relative amplitude of the states in each region depends on the
values of the electric and magnetic fields and thus it can be
tuned at will. For example, at a high magnetic field there are
states which peak very close to the centre of the well and decay
exponentially in the barrier. For low enough fields, Klein
tunneling is involved and the amplitude of the states in the
barrier increases drastically. Coupling between states which peak
in different regions results in the formation of anticrossing
points in the energy spectrum of the dot, provided the well is
deep. Numerical calculations suggest that the density of states is
relatively low between the few lowest Landau levels. Therefore, it
should be experimentally possible to probe individual quantum
states when the broadening of the Landau levels is small.

\section*{ACKNOWLEDGEMENTS}

We thank P.A. Maksym, Y. Bliokh, V. Freilikher and A. Rozhkov for
comments on the manuscript. G.G. acknowledges support from the
Japan Society for the Promotion of Science (JSPS) No.~P10502. F.N.
is partially supported by the ARO, NSF grant No.~0726909,
JSPS-RFBR contract No.~12-02-92100, Grant-in-Aid for Scientific
Research (S), MEXT Kakenhi on Quantum Cybernetics, and the JSPS
via its FIRST Program.


\begin{thebibliography}{10}
\bibitem{baregion}
The (electrostatic) barrier region is defined for $r>r_{0}$ and
the well for $r<r_{0}$, where $r_{0}$ satisfies $V(r_{0})=E$. The
actual region where the state is confined depends not only on the
external potential $V$ but also on the angular momentum and the
magnetic barrier due to the vector potential.
\bibitem{mneg}
For $m\le0$ there is an additional energy level which lies in the
gap.
\bibitem{coupling}
For some values of $m$ anticrossing points are also formed in the
upper energy ladder but not in the energy range of interest.
\bibitem{llevel}
The notation here implies that the Landau level $+1$ has energy
$+\sqrt{2e\hbar v^2_{F}B}$, the Landau level $-1$ has energy
$-\sqrt{2e\hbar v^2_{F}B}$, and so on for the higher levels.
\bibitem{peakregion}
If the magnetic field is low, the state in the lower energy ladder
peaks mainly in the barrier region and has a small amplitude in
the well region. When the field increases, the amplitude in the
well increases.
\end{thebibliography}
\end{document}